\documentclass[submission, Phys]{SciPost}

\def\C{{\mathcal{C}}}
\def\A{{\mathcal{A}}}
\def\F{{\mathcal{F}}}
\newcommand{\lb}{{\text{lb}}}
\newcommand{\ub}{{\text{ub}}}
\DeclareMathOperator{\tr}{tr}

\begin{document}

\begin{center}{\Large \textbf{
Quantum Many-body Bootstrap}}\end{center}

\begin{center}
X. Han\textsuperscript{1*}
\end{center}

\begin{center}
{\bf 1} Department of Physics, Stanford University, Stanford, CA 94305-4060, USA
\\
* hanxzh@stanford.edu
\end{center}

\begin{center}
\today
\end{center}


\section*{Abstract}
{\bf
A numerical bootstrap method is proposed to provide rigorous and nontrivial bounds in general quantum many-body systems with locality. In particular, lower bounds on ground state energies of local lattice systems are obtained by imposing positivity constraints on certain operator expectation values. Complemented with variational upper bounds, ground state observables are constrained to be within a narrow range. The method is demonstrated with the Hubbard model in one and two dimensions, and bounds on ground state double occupancy and magnetization are discussed.
}

\vspace{10pt}
\noindent\rule{\textwidth}{1pt}
\tableofcontents\thispagestyle{fancy}
\noindent\rule{\textwidth}{1pt}
\vspace{10pt}

\section{Introduction}
\label{sec:intro}
Understanding ground states of interacting many-body systems remains a central challenge in quantum physics. The general problem is intrinsically difficult \cite{complexity} and advances are often made with the aid of symmetries, approximations and numerics. Conformal symmetry and positivity have proved to be powerful in constraining correlators of quantum fields, via the conformal bootstrap \cite{RevModPhys.91.015002}. In this work the positivity constraints are applied to lattice systems without conformal invariance. 

The bootstrap approach in this work is algebraic in nature, and relies only on quantum mechanical first principles. As such it is capable of addressing ground state questions in systems with unbounded local Hilbert spaces, or with fermion sign problems.  For example, similar methods for solving many-body quantum mechanics with large-$N$ matrix degrees of freedom are proposed in \cite{Lin:2020mme, Han:2020bkb}. Also no approximation or assumption about the states is necessary, and thus the results are rigorous and serve as tests for other approximate algorithms. Generality and rigor of the method are favorable in cases where approximate methods give inconsistent results \cite{PhysRevX.5.041041}. 

The bootstrap algorithm is also a generalization of the established variational reduced density matrix theory \cite{rice2007reduced, densitymatrixtheory} to infinite lattices. In that method, the energy is minimized while the positivity constraints are imposed for few-body reduced density matrices, yielding lower bounds for ground state energies. Previous works (e.g., \cite{Baumgratz_2012, Verstichel2014, PhysRevLett.117.153001, doi:10.1063/1.5118899}) mostly deal with all two-body reduced density matrices, and hence the computational complexity is polynomial in system size. To better utilize geometric locality of the problem, I instead consider spatially local operators only. This allows me to systematically probe more-body operators and bootstrap directly in the thermodynamic limit. 

In this work some ground state observables in the Hubbard model \cite{doi:10.1098/rspa.1963.0204} are bounded, as a proof-of-principle demonstration of the method. In one dimension (see Table \ref{tab:e1}), exact solutions are available for comparison \cite{PhysRevLett.20.1445}. Significant numerical progress has been made in two-dimensional cases \cite{PhysRevX.5.041041, Zheng1155, Huang1161}. Lower bounds on ground state energies are obtained by bootstrap and are within a few percent of the state-of-the-art results (see Table \ref{tab:e2}). It should be interesting to compare the current algorithm with the Anderson bounds \cite{PhysRev.83.1260, PhysRevB.43.13743, PhysRevB.44.13203}. 

The lower bounds are complementary to the varitional upper bounds given by existing numerical approaches \cite{PhysRevX.5.041041}. Often the ground state energy and observables are then pinned down in a narrow range. Such rigorous constraints on ground state observables are not generally accessible to variational methods. As an example, nontrivial bounds on double occupancy and antiferromagnetic ordering in the two-dimensional Hubbard model ground states are obtained in Table \ref{tab:d} and \ref{tab:m}.

\section{Method}
The many-body bootstrap is based on symmetry and unitarity in quantum mechanics. Specifically, denote $\langle O \rangle = \tr (\rho O)$, where $\rho$ is some density matrix, then for any operator $O$, 
\begin{equation} \label{eq:cons}
    \langle I \rangle = 1, \quad \langle O^\dagger \rangle = \langle O \rangle^*, \quad \langle O^\dagger O \rangle \geq 0.
\end{equation}
Furthermore, $\langle U^{-1} O U \rangle = \langle O \rangle$ if $U$ is a symmetry of the state $\rho$, i.e., $U \rho U^{-1} = \rho$. If the symmetry is generated by a conserved charge $C$, also $ \langle [C, O] \rangle = 0$.
Thermal states and energy eigenstates are time translation invariant, so $\langle [H, O] \rangle = 0$ with $H$ the Hamiltonian and $O$ an arbitrary operator.

Lower bounds on ground state energies are obtained by minimizing $\langle H \rangle$ subject to the constraints (\ref{eq:cons}). More precisely, the minimization is done over the following set $\A$ of linear functionals $\F$ of operators:
\begin{align} \label{eq:full_cons}
    \A &= \{\F : \F[I] = 1, \quad \F[O^\dagger] = \F[O]^*,\nonumber\\
    &\quad \F[[C_\alpha, O]] = 0, \quad \F[U_\alpha^{-1} O U_\alpha] = \F[O], \quad \forall O \in \mathcal{C}_1, \nonumber \\
    &\quad \F[\widetilde{O}^\dagger \widetilde{O}] \geq 0, \quad \forall \widetilde{O} \in \mathcal{C}_2\}.
\end{align}
Minimization over this subset of functionals is equivalent to searching for operator expectation values $\langle O \rangle = \F[O]$ under the constraints (\ref{eq:cons}).
Here $C_\alpha$ and $U_\alpha$ are generators of the continuous and discrete symmetries to be imposed on the state. In practice the constraints (\ref{eq:cons}) can only be imposed for a subset of operators $\mathcal{C}_1$ and $\mathcal{C}_2$. Choice of $\C_1$ and $\C_2$ affects computational efficiency of the algorithm, and an empirical choice in fermionic lattice models will be discussed shortly.

The true ground state energy $E_0$ is bounded below by the minimal value from $\F \in \A$:
\begin{equation} \label{eq:lb}
    E_0 \geq \min_{\F \in \A} \F[H] =: E_{\lb},
\end{equation}
because the functional $\F[O] = \tr (\rho_0 O)$ is always in $\A$ for a ground state $\rho_0$ of $H$ that also commutes with all the charges $C_\alpha$ and $U_\alpha$. The minimization in (\ref{eq:lb}) can be solved efficiently and accurately by semidefinite programming (e.g., with \cite{ocpb:16, scs}). 

The equality in (\ref{eq:lb}) is reached when $\C_1$ and $\C_2$ are the full set of operators. Hence it is expected that the lower bound (\ref{eq:lb}) becomes tight as the number of constraints is increased. Indeed, any linear functional $\F$ can be written as $\F[O] = \tr (F O)$ for some operator $F$. And $F$ is a density matrix (positive with unit trace) if and only if (\ref{eq:cons}) holds for any $O$. Thus by the variational principle $\F[H] = \tr (F H)$ is minimized precisely when $F$ is a ground state, and $E_0 = \min \F[H]$. 

The bootstrap lower bound on ground state energy is complementary to the conventional variational upper bounds. Knowing that $E_{\text{lb}} \leq E_0 \leq E_{\text{ub}}$, the ground state expectation values can be bounded as
\begin{align} \label{eq:ob}
    \tr(\rho_0 O) &\geq \min_{\F \in \A, E_{\text{lb}} \leq \F[H] \leq E_{\text{ub}}} \F[O], \nonumber \\
    \tr(\rho_0 O) &\leq \max_{\F \in \A, E_{\text{lb}} \leq \F[H] \leq E_{\text{ub}}} \F[O].
\end{align}
The inequalities (\ref{eq:ob}) can be restrictive when $E_{\text{lb}}$ and $E_{\text{ub}}$ are close (e.g., see Table \ref{tab:e2} and \ref{tab:d}).

The method is illustrated with the Hubbard model in one and two dimensions:
\begin{equation} \label{eq:hubbard_h}
    H = - \sum_{\langle x y \rangle \sigma} c^{\dagger}_{x \sigma} c_{y \sigma} + U \sum_x n_{x \uparrow} n_{x \downarrow},
\end{equation}
where $\langle x y \rangle$ runs over ordered pairs of nearest-neighbor lattice sites, and $c_{x \sigma}$ is the fermion annihilation operator on site $x$ with spin $\sigma = \uparrow, \downarrow$. For simplicity I consider square lattices with unit spacing. The bootstrap  works directly in the thermodynamic limit. 

The Hamiltonian (\ref{eq:hubbard_h}) has discrete lattice translation and rotation symmetries, along with a U(2) global symmetry generated by
\begin{align} \label{eq:sym}
    N = \sum_x (n_{x \uparrow} + n_{x \downarrow}), \quad S_\alpha = \frac{1}{2} \sum_{x \sigma \sigma'} c_{x \sigma}^\dagger (\sigma_\alpha)_{\sigma \sigma'} c_{x \sigma'},
\end{align}
where $\alpha = x, y, z$ and $\sigma_\alpha$ are Pauli matrices. The fermion number $N$, total spin-$z$ component $S_z$, lattice translation and rotation will serve as $C_\alpha$ and $U_\alpha$ in (\ref{eq:full_cons}) for bootstrapping.

As mentioned previously, the choice of $\C_1$ and $\C_2$ in (\ref{eq:full_cons}) affects performance of the algorithm. In fermionic lattice models with a local Hamiltonian, such as (\ref{eq:hubbard_h}), it is plausible that local operators are more important. Dimensions of the subspaces $\C_1$ and $\C_2$ are controlled by a positive integer $K$, bounding the degree of locality of operators. The spaces are enlarged when $K$ is increased.

To be more precise, two types of locality are present in (\ref{eq:hubbard_h}): $k$-locality ($H$ is a sum of few-body operators) and geometric locality (the interactions are short-ranged). For a string of fermion creation and annihilation operators
\begin{equation} \label{eq:fermion_string}
    O = c_{x_1 \sigma_1}^{(\dagger)} c_{x_2 \sigma_2}^{(\dagger)} \ldots c_{x_r \sigma_r}^{(\dagger)},
\end{equation}
define a locality measure (with respect to a site chosen as the origin)
\begin{equation}
    l(O) = r + \sum_{i = 1}^r \|x_i\|.
\end{equation}
The first term $r$ is the number of fermion operators in (\ref{eq:fermion_string}), counting the degree of $k$-locality. The second term is a sum of geometric $l_1$-norms of the lattice vectors $x_i$. For any positive integer $K$, I choose $\C_2$ to be linearly spanned by fermion strings (\ref{eq:fermion_string}) with $l(O) \leq K$, and $\C_1$ spanned by the strings that appear in the products of two operators in $\C_2$. An ordering of fermion creation and annihilation operators is also employed and only normal ordered strings are considered to avoid unnecessary duplication. 

\begin{figure}[t]
    \centering
    \includegraphics[width=0.75\textwidth]{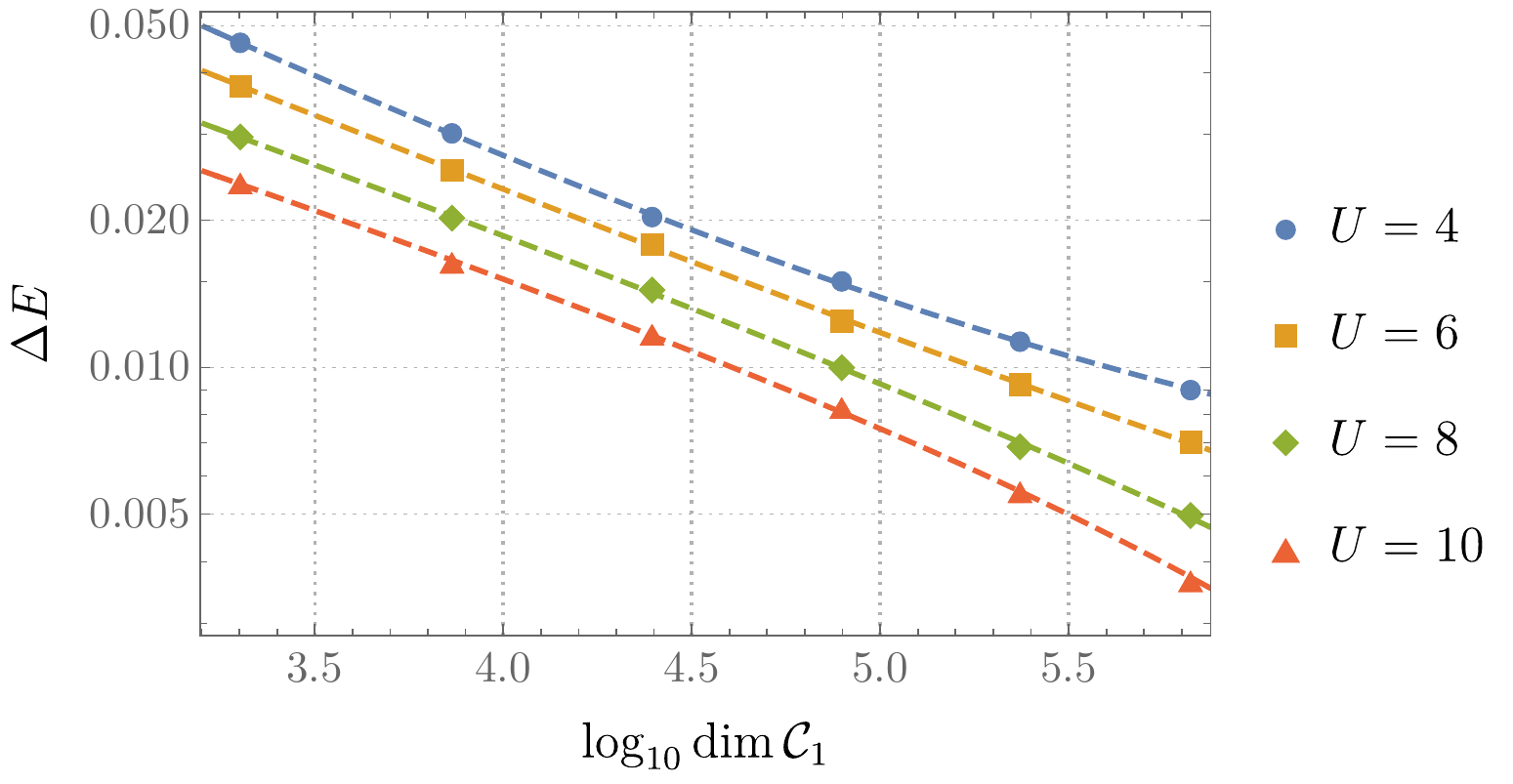}
    \caption{The difference $\Delta E = E_0 - E_\lb$ as a function of $\dim \C_1$, the number of operators in (\ref{eq:full_cons}), for the one-dimensional Hubbard model (\ref{eq:hubbard_h}) at half filling. Dashed curves show the best fits of form $E_\lb = A + B (\dim \C_1)^{- \alpha}$.}
    \label{fig:de}
\end{figure}

\begin{table}[t]
    \centering
    \begin{tabular}{||c|c|c|c|c||}
    \hline
        $n = 1$ & $U = 4$ & $U = 6$ & $U = 8$ & $U = 10$\\ \hline
        $E_\lb|_{K = 10}$     & $-0.5827$ & $-0.4271$ & $-0.3325$ & $-0.2708$ \\ 
        $E_\lb|_{K = \infty}$ & $-0.5781(7)$ & $-0.4212(9)$ & $-0.3260(11)$ & $-0.2648(14)$ \\ 
        $E_0$ & $-0.5737$ & $-0.4201$ & $-0.3275$ & $-0.2672$ \\ \hline
        $\F[D]|_{K = 10}$     & $0.1013$ & $0.0592$ & $0.0373$ & $0.0252$ \\ 
        $\F[D]|_{K = \infty}$ & $0.1015(4)$ & $0.0588(7)$ & $0.0371(4)$ & $0.0248(3)$ \\ 
        $\langle D \rangle_0$ & $0.1002$ & $0.0582$ & $0.0366$ & $0.0248$ \\ \hline
    \end{tabular}
    \caption{Bootstrap lower bounds $E_\lb$ of one-dimensional Hubbard model ground state energies (per site), and the double occupancy $D$ in $\F$ that minimizes (\ref{eq:lb}). Exact values $E_0$ and $\langle D\rangle_0$ are shown for comparison. The number of fermions per site $n = 1$. For values extrapolated to $K = \infty$, standard errors in fitting are shown in the brackets. }
    \label{tab:e1}
\end{table}

\section{Result in Hubbard model}

\subsection{One dimension}
Symmetries imposed in (\ref{eq:full_cons}) include $C_\alpha = \{H, N, S_z\}$ from (\ref{eq:hubbard_h}) and (\ref{eq:sym}), and $U_\alpha = \{T, \Pi\}$. Here $T$ is the lattice translation and $\Pi$ the lattice reflection. For $5 \leq K \leq 10$, $E_{\lb}$ in (\ref{eq:lb}) is evaluated and lower bounds the ground state energy. The best bound from $K = 10$ is shown in Table \ref{tab:e1}. Other expectation values are also available, for the functional $\F$ that minimizes (\ref{eq:lb}). For example, $D = n_{x \uparrow} n_{x \downarrow}$ in Table \ref{tab:e1} is the double occupancy. Note that $\F[D]$ does not necessarily bound the ground state value $\langle D \rangle_0 = \tr (\rho_0 D)$. 

Extrapolation to $K = \infty$ is also possible. In Figure \ref{fig:de} expectation values at finite $K$ fit well to the functional form $A + B (\dim \C_1)^{-\alpha}$, where $\dim \C_1$ is the number of operators in the constraints (\ref{eq:full_cons}). The fitted $\alpha \approx 0.3$, consistent with that the algorithmic complexity is polynomial in the required accuracy. Standard errors from the fitting are included in Table \ref{tab:e1}. The extrapolated values agree with the exact solution. 

\begin{table}[t]
    \centering
    \begin{tabular}{||c|c|c|c|c||}
    \hline
        $n = 1$ & $U = 2$ & $U = 4$ & $U = 6$ & $U = 8$\\ \hline
        $E_\lb|_{K = 7}$     & $-1.221$ & $-0.913$ & $-0.705$ & $-0.565$ \\ 
        $E_\lb|_{K = \infty}$ & -- & -- & $-0.66(2)$ & $-0.54(2)$ \\ 
        $E_{\text{AFQMC}}$ & $-1.1763(2)$ & $-0.8603(2)$ & $-0.6568(3)$ & $-0.5247(2)$ \\
        $E_{\text{DMET}}$ & $-1.1764(3)$ & $-0.8604(3)$ & $-0.6562(5)$ & $-0.5234(10)$ \\ 
        $E_{\text{DMRG}}$ & $-1.176(1)$ & $-0.8605(5)$ & $-0.6565(1)$ & $-0.5241(1)$ \\
        \hline
        $n = 0.875$ & $U = 2$ & $U = 4$ & $U = 6$ & $U = 8$\\ \hline
        $E_\lb|_{K = 7}$     & $-1.316$ & $-1.103$ & $-0.963$ & $-0.867$ \\ 
        $E_\lb|_{K = \infty}$ & -- & -- & $-0.86(5)$ & $-0.77(3)$ \\ 
        $E_{\text{DMET}}$ & $-1.2721(6)$ & $-1.031(3)$ & $-0.863(13)$ & $-0.749(7)$ \\ \hline
    \end{tabular}
    \caption{Bootstrap lower bounds $E_\lb$ of two-dimensional Hubbard model ground state energies (per site) $E_0$, at fillings $n = 1$ and $n = 0.875$. Solutions from AFQMC, DMET and DMRG are shown for comparison.}
    \label{tab:e2}
\end{table}

\subsection{Two dimensions}
Symmetries are $C_\alpha = \{H, N, S_z\}$ along with $U_\alpha = \{T_{(1, 0)}, T_{(0, 1)}, \Pi, R\}$, where $T_{(1, 0)}$ and $T_{(0, 1)}$ are the lattice translations, $\Pi$ the reflection, and $R$ the $\pi/2$ lattice rotation. No exact solution is known for general couplings, and I will compare with the AFQMC \cite{zhang201315}, DMET \cite{PhysRevLett.109.186404, PhysRevB.93.035126} and DMRG \cite{PhysRevLett.69.2863} results reviewed in \cite{PhysRevX.5.041041}. Here the AFQMC solution is numerically exact at half filling without sign problems. The DMRG is a variational technique and the DMET is not variational. 

The bounds from $K = 7$, along with the values extrapolated  to $K = \infty$ from $4 \leq K \leq 7$, are obtained in Table \ref{tab:e2}, \ref{tab:d} and \ref{tab:m}. Estimated standard errors are shown in the brackets. Some values are omitted due to deficient $K$ and thus poor fitting quality in extrapolation. While the bounds are rigorous for any finite $K$, uncontrolled errors are introduced in extrapolation. The extrapolation may be further improved with more computational resources.

For smaller $U$ in Table \ref{tab:e2}, the best bounds available are within a few percent of the variational energies, corroborating the effectiveness of both methods. At larger $U$, when extrapolation is more reliable, the extrapolated energies agree with other numerics within numerical uncertainties.

\begin{table}[t]
    \centering
    \begin{tabular}{||c|c|c|c|c||}
    \hline
        $n = 1$ & $U = 2$ & $U = 4$ & $U = 6$ & $U = 8$\\ \hline
        $d_\lb|_{K = 7}$ & $0.160$ & $0.106$ & $0.071$ & $0.049$ \\ 
        $d_\lb|_{K = \infty}$ & $0.161(6)$ & $0.108(7)$ & $0.072(5)$ & $0.050(3)$ \\ 
        $d_\ub|_{K = 7}$ & $0.224$ & $0.169$ & $0.117$ & $0.079$ \\ 
        $d_\ub|_{K = \infty}$ & $0.195(14)$ & -- & -- & -- \\ 
        $d_{\text{DMET}}$ & $0.1913(4)$ & $0.1261(1)$ & $0.08095(4)$ & $0.05398(7)$ \\ 
        $d_{\text{DMRG}}$ & $0.188(1)$ & $0.126(1)$ & $0.0809(3)$ & $0.0539(1)$ \\
        \hline
    \end{tabular}
    \caption{Bootstrap bounds $d_\lb \leq \tr (\rho_0 D)  \leq d_\ub$ of ground state double occupancy (per site) $D = n_{x \uparrow} n_{x \downarrow}$, for the two-dimensional Hubbard model at half filling.}
    \label{tab:d}
\end{table}

\begin{table}[t]
    \centering
    \begin{tabular}{||c|c|c|c|c||}
    \hline
        $n = 1$ & $U = 2$ & $U = 4$ & $U = 6$ & $U = 8$\\ \hline
        $m_\ub|_{K = 7}$     & $0.194$ & $0.292$ & $0.352$ & $0.383$  \\ 
        $m_\ub|_{K = \infty}$ & -- & -- & -- & $0.34(2)$ \\ 
        $m_{\text{DMET}}$ & $0.133(5)$ & $0.252(9)$ & $0.299(12)$ & $0.318(13)$ \\ \hline
    \end{tabular}
    \caption{Bootstrap upper bounds $m_\ub$ of ground state staggered magnetization (\ref{eq:mag}) per site, at half filling.}
    \label{tab:m}
\end{table}

If the ground state energies in \cite{PhysRevX.5.041041} are upper bounds, local observables are constrained by (\ref{eq:ob}). For instance, in the following the DMRG energies at $n = 1$ from \cite{PhysRevX.5.041041} are used as $E_\ub$. Bounds for double occupancy $D$ are shown in Table \ref{tab:d}, which are restrictive and consistent with other numerics.

As another example, consider the staggered magnetization
\begin{equation} \label{eq:mag}
    M = \frac{1}{2} \sum_{x} (-1)^{x_1 + x_2} (n_{x \uparrow} - n_{x \downarrow}),
\end{equation}
where $(x_1, x_2)$ are coordinates of $x$. The set of discrete symmetries is reduced to $U_\alpha = \{T_{(1, 1)}, T_{(1, -1)}, \Pi, R\}$, to allow for nonzero $M$. Upper bounds on $M$ per site are obtained in Table \ref{tab:m}. At large $U$ the bound is also consistent with the Heisenberg limit $m \approx 0.307$ \cite{PhysRevB.56.11678}. For magnetization the two inequalities in (\ref{eq:ob}) are not independent, as $\min \F[M] = - \max \F[M]$.

\section{Conclusion}
I have shown that the idea of positivity, which is fundamental in many successful theories, can be employed to solve local lattice models. The bounds are nontrivial checks on other numerics and expand our knowledge of interacting quantum many-body systems. 

It would be ideal to have a nonzero lower bound on ground state ordering as well. This is difficult in the current formalism as ground states that do not break symmetries are not ruled out by the constraints. Possibly one should consider two-point functions, by re-introducing non-local few-body operators of interest. 

Other directions include generalizing the method to continuous theories, or imposing more constraints on the state (for example, that the state is thermal or a condensate). Also bootstrap bounds on spectral functions, as well as inhomogeneous phases may be useful in constraining low-energy excitations and competing orders in strongly correlated electron systems.

\section*{Acknowledgements}
This work arose from discussions with Sean Hartnoll. The author acknowledges discussions with Sean Hartnoll, Jorrit Kruthoff, Edward Mazenc and Daniel Ranard on related projects.



\bibliography{boot.bib}

\nolinenumbers

\end{document}